# Angle-resolved polarized Raman spectroscopy for distinguishing stage-I graphite intercalation compounds with Thorium, Uranium and Plutonium

Kun Yan[1 a)]

[1]State Key Laboratory of Optoelectronic Materials and Technologies, Guangdong Province Key Laboratory of Display Material and Technology, School of Physics, Sun Yat-sen University, Guangzhou 510275, People's Republic of China.

[a)]Correspondence: yank6@mail2.sysu.edu.cn

## Abstract

Graphite intercalation compounds (GICs) with the geometrical anisotropy and strong electron-phonon coupling are in full swing and have shown their great potential for applications in nanodevices. I selected representative three elements in actinide group with valence electron arrangement: Thorium (Th) ([Rn]$6d^2 7s^2$), Uranium (U) ([Rn]$5f^3 6d^1 7s^2$), Plutonium (Pu) ([Rn]$5f^6 7s^2$). I calculated their phonon spectra and demonstrated the atomic-scale microstructure identification of actinide graphite intercalation compounds by angle-resolved polarized Raman spectroscopy.

# 1  |  Introduction

The synthesis of a graphite intercalation compound (GIC) was first reported by Schaffäutl (1841). Due to its unique electronic structure and Fermi surface, transport characteristics, optical spectra, magneto-optical phenomenon, magnetic susceptibility and magnetic resonance, specific heat, superconductivity, high conductivity ($5.8*10^5$ $(\Omega$ cm$)^{-1}$, comparable to copper), lattice vibration modes, the GICs' potential is unlimited in the application of battery and electrodes, catalysis, , laser diodes, high-power devices, sensors, etc. Alkali and alkaline earth metals have been artificially intercalated into graphite sheets to achieve superconducting properties[1-3]. GICs of transition metal elements have also been partially studied, but due to the complexity of molecular orbitals and electronic structures, there has been no generally accepted and consistent conclusions/rules to be achieved. The GICs with lanthanide (Ln) and actinide (An) groups containing f-orbital electrons are even less studied. So far, only the similar $\eta^6$-benzene or $\eta^8$-cyclooctatetraene (COT) Ln and An complexes have been chemically synthesized[4]. Nevertheless, the Ln and An groups include extremely important elements, such as Thorium (Th), Uranium (U), Plutonium (Pu) and other raw materials for nuclear energy utilization. Furthery, Th ($[Rn]6d^27s^2$) has valence electrons of only d and s orbit, Pu ($[Rn]5f^67s^2$) has (ones of) only f and s orbit, and U ($[Rn]5f^36d^17s^2$) has all.

Angle-resolved Photo-emission Spectroscopy (ARPES) is usually employed to explore the electronic properties of the GICs with alkali and alkaline earth metal elements. However, due to the anisotropy and strong electron-phonon coupling (EPC), as has been discovered, Angle-resolved Polarized Raman Spectroscopy (APRS) may be a fast, effective, and nondestructive alternative to fingerprint differences at the atomic scale within similar molecular structures[5-7].

In this work, I calculated the Phonon dispersion and Angle-resolved Polarized Raman Intensity (APRI) of stage-I GICs with Th, U and Pu. I discovered that the GICs are all dynamically stable. Thus, I further simulated the APRI of the six types with different metal elements and their various densities and distribution features. It is demonstrated that the sharp distinctions in APRI can be a clear and easy basis for identification of geometry and the crystallographic orientation.

## 2 | Model and Methods

Six types of stage-I GICs with actinide (An) group metal elements Thorium (Th), Uranium (U) and Plutonium (Pu) were designed. 'stage-I' was experimentally reported[8] to have the strongest conductivity, just like monolayer graphene to which metal nanoparticles attach. It is explained by the strong EPC effect. The metal atoms are so large that 2-dimentianal carbon layer is basically tiled by octagons and quadrilaterals[9] (see Fig. 1), and the metal atoms bond with carbons in the upper and lower octagons. After geometric optimization, the C-C bond length between two adjacent octagons is 1.44 Å; the C-C bond length between the quadrate and octagon is 1.48 Å. Those C-C bond lengths are consistent with previous research results[9], which notes the former as "short" ones and the latter as "long" ones. The distance between the upper and lower adjacent carbon layers ranges from 3.8 to 4.3 Å, according to the metal element and its density. When the metal atoms fill the middle of the octagon (so its density marks "100%"), such as 2DU100, the distance between adjacent U atoms is 3.45 Å. When the density of metal atoms is "50%", as shown in Fig. 1(c), the distance between adjacent U atoms is 4.87 Å. Fig. 1(e) and (f) indicate 2 way of metal atom (e.g. U) distribution with the same density ("25%"), noted as 2DU25-1 and 2DU25-2. As I are focus on demonstrating the probability and reliability of APRI's fingerprint, no more discussion of else density and distribution way is to be added in this paper.

Geometric optimization and electronic structure were carried out with density functional theory (DFT) as implemented in Vienna Ab initio simulation package (VASP) code. The exchange-correlation potential was treated at the level of generalized gradient approximation (GGA) using Perdew–Burke–Ernzerhof (PBE) functionals. A plane-wave energy cutoff of 480 eV was employed in all calculations. The force and electronic convergence tolerance were set to $10^{-5}$ eV/Å and $10^{-6}$ eV, respectively in the full geometrical optimizations. The k-point grid was set to 8×8×4 in structure optimization. Spin polarization was performed and due to the localized $d$- and $f$-orbital electrons a GGA+U approach was used in all calculations. The vibrational frequencies and eigenvectors of phonons were obtained with Phonopy code [10]. 2×2×2 supercells were chosen in phonon calculations.

The Raman scattering cross section of the Stokes part per unit cell is [11-19]

$$\frac{d\sigma_\upsilon}{d\Omega} = V_0 \frac{(\omega_i - \omega_\upsilon)^4}{(4\pi)^2 c^4} \left| \vec{e}_s \cdot \vec{R}_\upsilon \cdot \vec{e}_i \right|^2 (n_\upsilon + 1) \frac{\hbar}{2\omega_\upsilon} \qquad (1)$$

$$\vec{R}_{\alpha\beta}(\upsilon) = \sqrt{V_0} \sum_{\mu=1}^{N} \sum_{l=1}^{3} \frac{\partial \chi_{\alpha\beta}}{\partial r_l(\mu)} \frac{e_l^\upsilon(\mu)}{\sqrt{M_\mu}} \qquad (2)$$

where $V_0$ is the unit cell volume; $\omega_i$ and $\omega_\upsilon$ are the frequencies of the incident light and the $\upsilon$-th vibrational mode, respectively; $\vec{e}_s$ and $\vec{e}_i$ are the polarization directions of the scattered light and the incident light, respectively; $\vec{R}_\upsilon$ is the Raman tensor; $n_\upsilon$ is the Bose factor; $\frac{\partial \chi_{\alpha\beta}}{\partial r_l(\mu)}$ is the first-order derivative of the frequency-dependent linear dielectric susceptibility tensor $\chi$ with respect to $r_l(\mu)$ (the displacement of the $\mu$-th atom along direction $l$); $e_l^\upsilon(\mu)$ is the eigenvector of the $\upsilon$-th phonon mode at Γ point; $c$ is the light velocity in vacuum; $\hbar$ is Reduced Planck Constant; $M_\mu$ is the molecular weight of the $\mu$th atom.

## 3 | Results and discussion

### 3.1 | Phonon Dispersion

No imaginary value in the Phonon Dispersion means that all the six types of GICs have dynamic stability. That the spectra of 2DU25-1 and 2DU25-2 have more denser curves is because their unit cells have twice as many atoms as the other structures. The metal atoms are far larger and heavier than C atoms, so their projected density of states (PDOS) contribute to low frequency phonons.

### 3.2 | Angle-resolved polarized Raman spectroscopy (APRS) for distinction

Our calculation coincides with group theory (shown in Table 1). Thus, the Raman tensor can be written in a general form:

$$R = \begin{bmatrix} a & d & e \\ d & b & f \\ e & f & c \end{bmatrix} \quad (3)$$

The intensity of APRS (APRI) is calculated in a backscattering configuration, while the polarization of the scattered light can be perpendicular ($\perp$) or parallel ($//$) to that of the incident light. When the *k* vector of the incident laser is in y direction ($\hat{k}_i \parallel y$) with polarization direction $\vec{e}_i = (\cos\theta, 0, \sin\theta)$, the polarization directions of scattered light in parallel and perpendicular configuration can be written as $\vec{e}_s^{\parallel} = (\cos\theta, 0, \sin\theta)$ and $\vec{e}_s^{\perp} = (-\sin\theta, 0, \cos\theta)$, respectively. That way can be used to study the interaction and EPC between the metal atoms and the carbon 2-dimentional network. When the *k* vector of the incident laser is in z direction ($\hat{k}_i \parallel z$) with polarization direction $\vec{e}_i = (\cos\theta, \sin\theta, 0)$, the polarization directions of scattered light in parallel and perpendicular configuration can be written as $\vec{e}_s^{\parallel} = (\cos\theta, \sin\theta, 0)$ and $\vec{e}_s^{\perp} = (-\sin\theta, \cos\theta, 0)$, respectively. That way can study the anisotropy in xy plane, and the various distribution feature of metal atoms.

Usually, in experiments we know the z-axis direction of the sample (using an optical microscope would be ok). Since the angle between the polarization direction of the incident light and the x-axis direction is set to an arbitrary $\theta$, the x and y axis directions could be set arbitrary, just required to be in a plane perpendicular to the z axis. Thus we set the laboratory coordinates as shown in Figure 1(a) and (g).

Therefore, by substituting into formula (2) the above $\vec{e}_i$, $\vec{e}_s$ and Raman tensor $\vec{R}$ calculated with VASP according to formula (1), I have the intensity for the parallel-polarization configuration 1 when $\hat{k}_i \parallel y$:

$$I_{\hat{k}_i\parallel y}^{\parallel}(\theta) \propto \left| a\cdot\cos^2\theta + c\cdot\sin^2\theta + e\cdot\sin 2\theta \right|^2 \cdot \frac{(n_v + 1)\cdot\omega_i^4}{\omega_v} \quad (4)$$

and the cross-polarization configuration 2 when $\hat{k}_i \parallel y$:

$$I_{\hat{k}_i\parallel y}^{\perp}(\theta) \propto \left| \frac{(c-a)}{2}\cdot\sin 2\theta + e\cdot\cos 2\theta \right|^2 \cdot \frac{(n_v + 1)\cdot\omega_i^4}{\omega_v} \quad (5)$$

Similarly, I have the intensity for the parallel-polarization configuration 3 when $\hat{k}_i \parallel z$:

$$I^{\parallel}_{\hat{k}_i \parallel z}(\theta) \propto \left| a \cdot \cos^2 \theta + b \cdot \sin^2 \theta + d \cdot \sin 2\theta \right|^2 \cdot \frac{(n_\upsilon + 1) \cdot \omega_i^4}{\omega_\upsilon} \tag{6}$$

and the cross-polarization configuration 4 when $\hat{k}_i \parallel z$:

$$I^{\perp}_{\hat{k}_i \parallel z}(\theta) \propto \left| \frac{(b-a)}{2} \cdot \sin 2\theta + d \cdot \cos 2\theta \right|^2 \cdot \frac{(n_\upsilon + 1) \cdot \omega_i^4}{\omega_\upsilon} \tag{7}$$

Based on formulas (4) to (7), we obtained the APRI of the prominent peaks of the GICs (Fig. 2-5). For 2DTh50, when $\hat{k}_i \parallel y$, there is a strong peak of 404 cm$^{-1}$ mode for A$_g$ in configuration 1 and 2, and an obvious curve rotation of 307.5 cm$^{-1}$ due to the non-negligible absorption of 633 nm light; due to $|a|=|b|<|d|$, 1119.4 cm$^{-1}$ mode has a particular shape in configuration 3 and a four-fold symmetry with axis $\theta=0°$ and $\theta=90°$ in configuration 4. For B$_g$, the 1285.9 cm$^{-1}$ mode is so strong that it can be used to identify such structure. Also, the shape of the mode is like an ellipse with the major axis twice the minor axis, due to $|a|\approx|b|\approx 2|d|$. Similar analysis can be applied to other structures. For 2DPu50, the 1098.3 cm$^{-1}$ mode and the 1099.9 cm$^{-1}$ mode are highly strong for identification in A$_g$. For 2DU50, the strong 1447 cm$^{-1}$ mode has four-fold symmetry with axis $\theta=0°$ and $\theta=90°$ in configuration 4 and a "asymmetric four petals" shape (due to real part: $|a|\approx|b|\approx|d|$, imaginal part $\approx$ 1/10*real part ) in configuration3. Also, the 306.4 cm$^{-1}$ mode has special shapes in both configuration 1 and 2 (due to $|a|\approx|e|\gg|c|$). Both features can be used to identify 2DU50. If $|a|\approx|b|\gg|d|$, the shape is a circle and if $|a|\neq|b|\gg|d|$, an ellipse. If $|d|\neq 0$, we attain a "spindle body" with $I\neq 0$ when $\theta=90°$, which is the prominent feature for 2DU100 in configuration 1. There is no obvious rotation of APRI curves in 2DU100, due to the possibly weak absorption of the relatively low frequencies of selected phonons. The 581.5 cm$^{-1}$ mode and 360.3 cm$^{-1}$ mode in configuration 1 and 452.2 cm$^{-1}$ mode in configuration 3 can be used to identify 2DU25_1 due to their peculiar shape. The appearance of dense strong peaks with various rotation in configuration 4 can be another basis for identification. For 2DU25_2, in configuration 3 the curves all have an oval shape and in configuration 4 A$_{1g}$ modes and B$_{2g}$ modes have four-fold symmetry with axis $\theta=0°$ and $\theta=90°$, which both are obviously different from 2DU25_1. I consider those can be used to distinguish 2DU25_1 and 2DU25_2.

Moreover, comparing the experimental results with the theoretical simulation above, I can easily confirm the configuration and furtherly the x/y/z-axis's direction. For example, the $\theta_0$ can be determined by $\left.\frac{dI}{d\theta}\right|_{\theta_0}=0$, where $I$ takes extremum when $\theta=\theta_0$. Then since the x or y-axis' direction has been set to along $\theta=0°$ or $\theta=90°$ respectively, the crystallographic orientation can be determined, e.g. x axis, by rotating clockwise $\theta_0$ the angle of symmetry axis of the APR curves based on our calculation.

### 3.3 | Some problems in experiment

Owing to the big size of the actinide metal atoms, the preparing method diffusing alkali and alkaline earth metals into graphite from the edge of the sheets is never applied. There are 2 probable routes for the preparation of the GICs in this work:

1. Using the means of thin film growth: By MOCVD, etc., add a layer of actinide metal atoms on the top of a layer of graphene, then cover the material with another layer of graphene, and then grow another metal layer...Thus, metal atomic layers and graphene are alternately stacked and grown layer by layer.

2. Using the hydrostatic pressure: The high pressure drives the metal atoms slowly pass through the carbon ring with relaxed structure and gradually enter the inner of graphite.

Like high-density solid-state hydrogen storage materials, reducing pressure or heating could release metals from GICs. Also, those GICs could be expected to be recycled.

There are other questions: what is the resolution needed to separate to neighboring lines? When a focusing lens is used to have a micron size laser beam, the laser beam is no more completely perpendicular to the surface; hence, zero Raman intensity cannot be observed: how does this effect modify the calculated Raman curve? Owing to the sample not yet prepared, the lack of comparison between calculated curves and experimental results is a serious drawback. The author find a literature entitled "The high-resolution Raman spectroscopy of crystalline uranocene, thorocene, and ferrocene"[20]. I calculated the phonon spectra of those crystals again and found the imaginary frequency disappeared when I selected a correct/right U value (see Fig. S1). Also, the obtained frequencies are closer to the experimental (RUN) data than the DFT results in the literature (see Table S1). The deviation may stem/come from the ignorance of vdW correction.

## 4 | Conclusions

Graphite intercalation compounds (GICs) with unique physical properties are in full swing and have shown their great potential for applications in nanodevices. I selected representative three elements of/in actinide group with valence electron arrangement: Thorium (Th) ([Rn]$6d^27s^2$), Uranium (U) ([Rn]$5f^36d^17s^2$), Plutonium (Pu) ([Rn]$5f^67s^2$). I calculated their phonon spectra and demonstrated the atomic-scale microstructure identification of actinide graphite intercalation compounds by APRI. The phonon dispersions of their stage-I graphite intercalation compounds (GICs) were calculated and they are all found to be dynamically stable. Further, I simulated the APRI of those GICs with different densities and distribution way, which demonstrated that the sharp distinctions in APRI can be a clear and easy basis for identification of geometry and the crystallographic orientation.

**ACKNOWLEDGEMENTS**

The author thanks the Physical Research Platform (PRP) in School of Physics, SYSU.

**DATA AVAILABILITY STATEMENT**

The data that support the findings of this study are available from the corresponding author upon reasonable request.

# NOTES AND REFERENCES

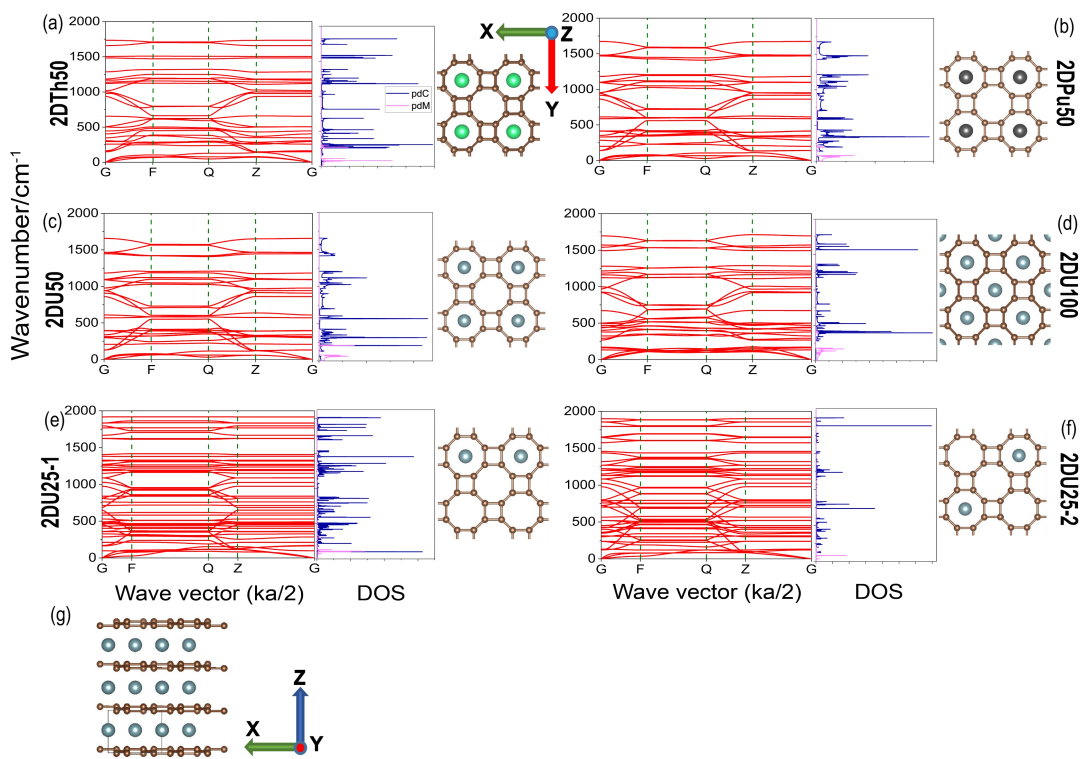

Fig.1 Phonon dispersion (left panels), phonon projected density of states (PDOS) (middle panels) and structural diagram viewed along Z-axis (right panels) of GICs with (a) Th (50%), (b) Pu (50%), (c) U (50%), (d) U (100%), (e) U (25%-1) and (f) U (25%-2). In the PDOS panels, pink curves are PDOS of metal elements and blue ones are that of carbons. In the structural diagrams, green balls are Th atoms, black ones are Pu, blue ones are U and brown ones are carbon. (g) structural diagram of 2DU100 viewed along Y-axis.

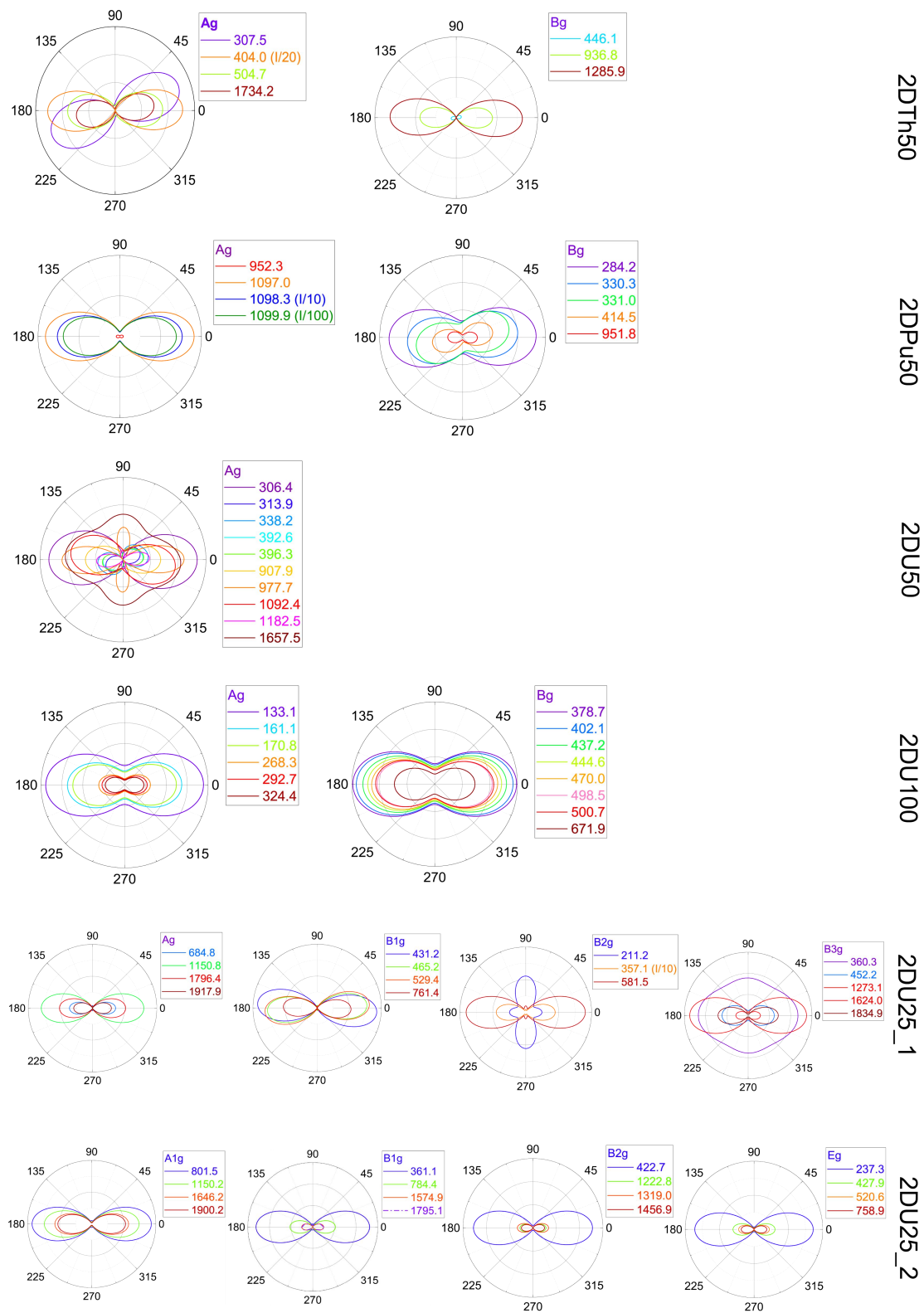

Fig. 2 Angle-resolved polarized Raman spectroscopy (APRS) of 6 types GICs with 633 nm laser incident along y-axis ($\hat{k}_i \parallel y$). The scatted light employs a parallel-polarization configuration ($\vec{e}_s \parallel \vec{e}_i$).

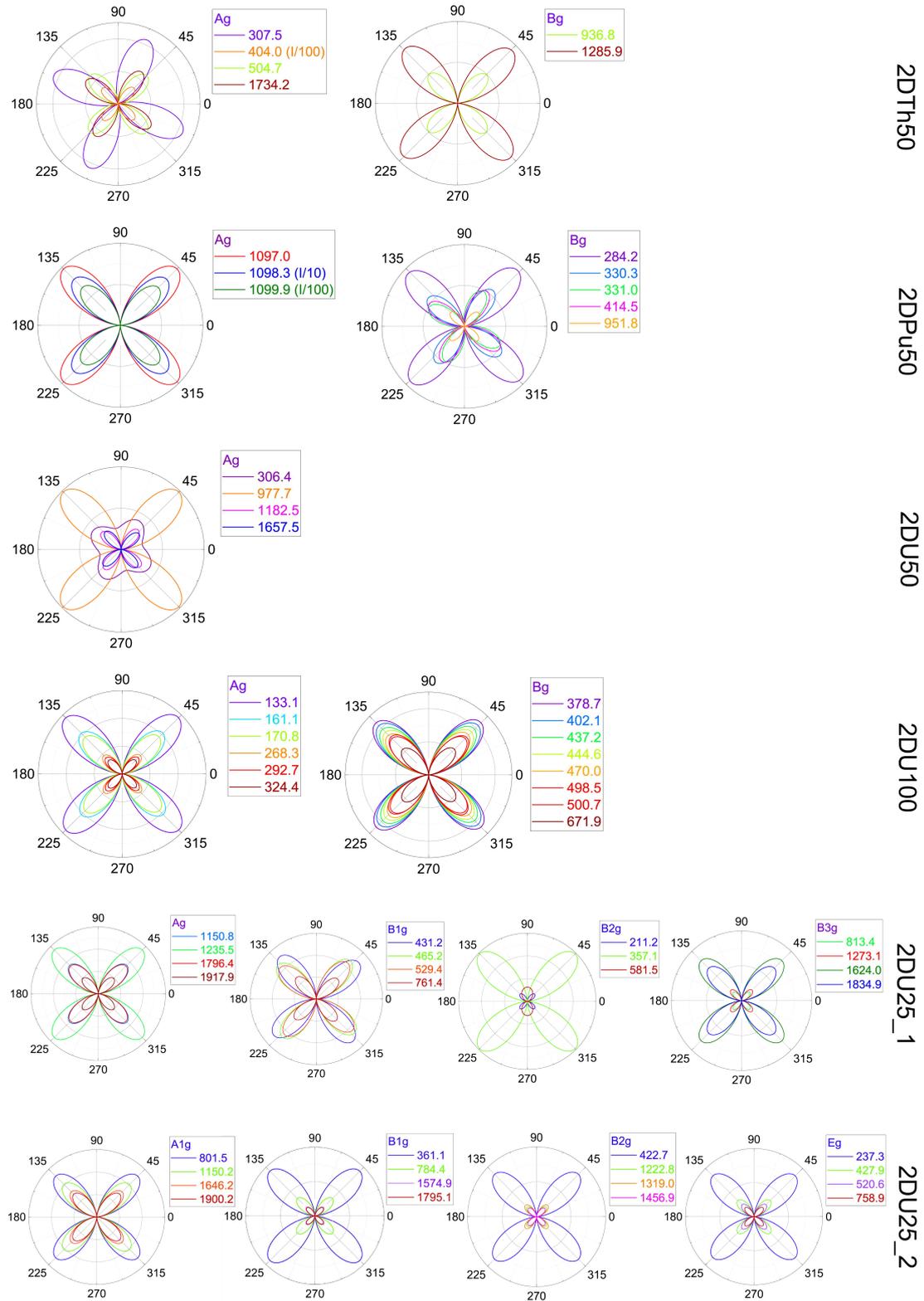

Fig. 3 Angle-resolved polarized Raman spectroscopy (APRS) of 6 types GICs with 633 nm laser incident along y-axis ($\hat{k}_i \parallel y$). The scatted light employs a cross-polarization configuration ($\vec{e}_s \perp \vec{e}_i$).

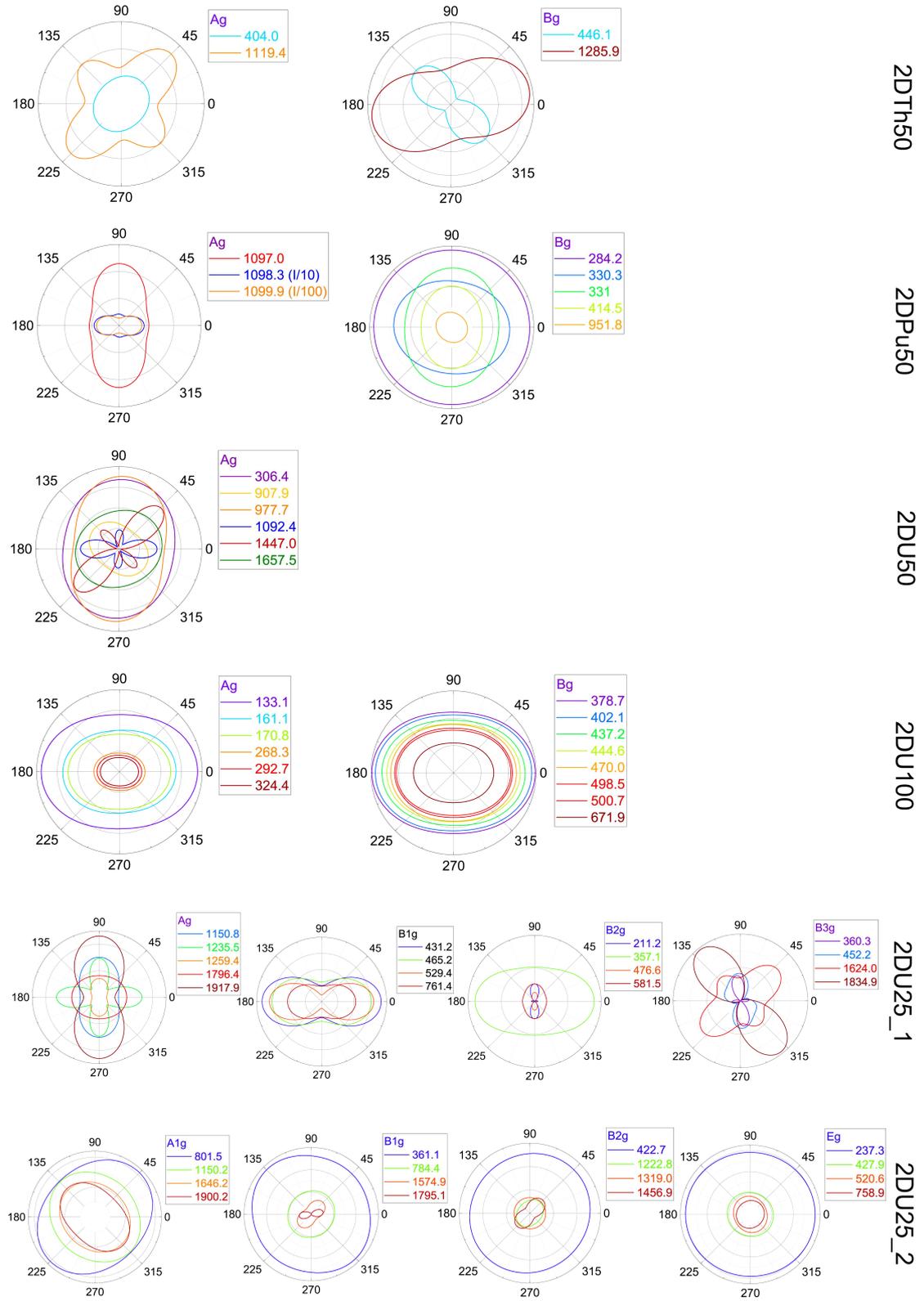

Fig. 4 Angle-resolved polarized Raman spectroscopy (APRS) of 6 types GICs with 633 nm laser incident along y-axis ($\hat{k}_i \parallel z$). The scatted light employs a parallel-polarization configuration ($\vec{e}_s \parallel \vec{e}_i$).

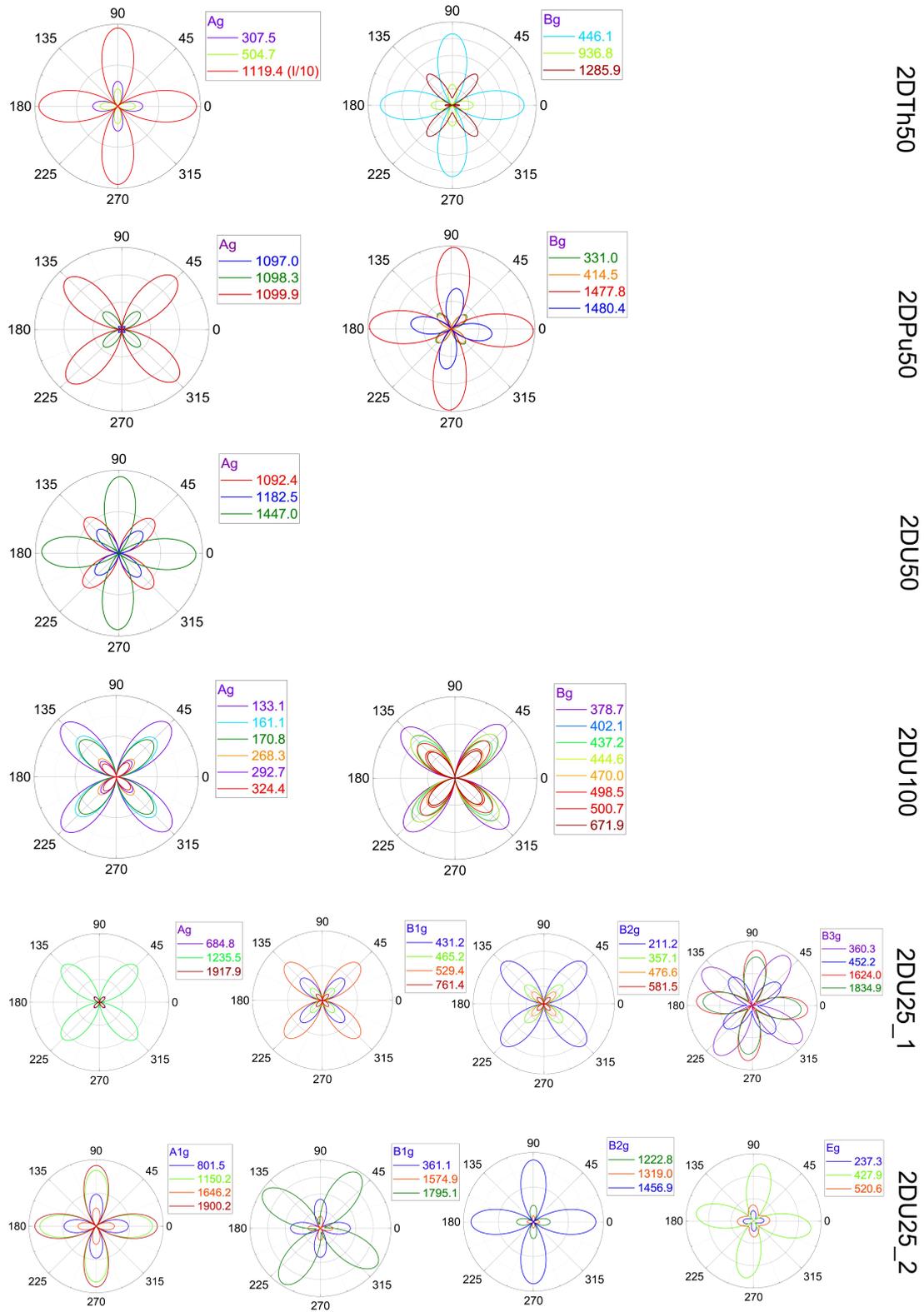

Fig. 5 Angle-resolved polarized Raman spectroscopy (APRS) of 6 types GICs with 633 nm laser incident along y-axis ($\hat{k}_i \parallel z$). The scatted light employs a parallel-polarization configuration ($\vec{e}_s \perp \vec{e}_i$).

Table 1. The Point group, Raman active modes and relative functions of the designed Organometallics

| OM name | Point Group | Raman Active Modes | functions |
|---|---|---|---|
| 2DUTh50 | $C_{2h}$ (2/m) | $A_g$ | $x^2, y^2, z^2, xy, Jz$ |
| | | $B_g$ | $xz, yz, Jx, Jy$ |
| 2DPu50 | $D_{4h}$ (4/mmm) | $A_{1g}$ | $x^2+y^2, z^2$ |
| | | $B_{1g}$ | $x^2-y^2$ |
| 2DU50 | $D_{4h}$ (4/mmm) | $A_{1g}$ | $x^2+y^2, z^2$ |
| 2DU100 | $C_{2h}$ (2/m) | $A_g$ | $x^2, y^2, z^2, xy, Jz$ |
| | | $B_g$ | $xz, yz, Jx, Jy$ |
| 2DU25_1 | $D_{2h}$ (mmm) | $A_g$ | $x^2, y^2, z^2$ |
| | | $B_{1g}$ | $xy, Jz$ |
| | | $B_{2g}$ | $xz, Jy$ |
| | | $B_{3g}$ | $yz, Jx$ |
| 2DU25_2 | $D_{4h}$ (4/mmm) | $A_{1g}$ | $x^2+y^2, z^2$ |
| | | $B_{1g}$ | $x^2-y^2$, |
| | | $B_{2g}$ | $xy$ |
| | | $E_g$ | $(xz, yz), (Jx, Jy)$ |

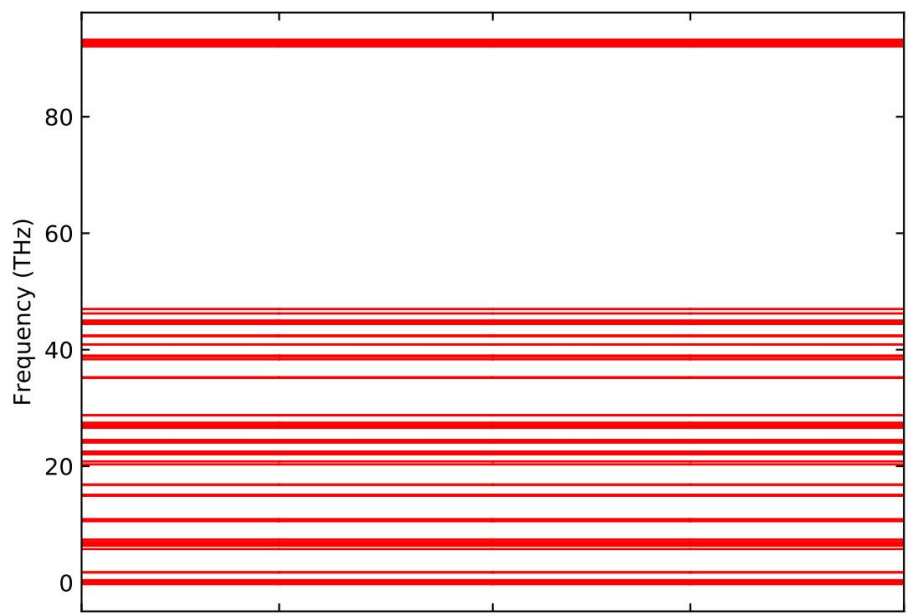

Fig. S1 Phonon spectrum of uranocene (calculated by GGA+U).

Table S1 Phonon frequencies calculated by GGA+U in this work (right table). They are closer to the experimental (RUN) data than the DFT results in the literature[20] (left table). The deviation may stem/come from the ignorance of vdW correction.

TABLE II. The observed RUN and calculated DFT Raman frequencies of $Th(C_8H_8)_2$, $U(C_8H_8)_2$, and $C_8H_8^{2-}$ ion with $D_{8h}$ symmetry.

Results of this work

| Mode | $C_8H_8^{2-}$ | $U(C_8H_8)_2$ | | $Th(C_8H_8)_2$ | | $U(C8H8)2$ | | $Th(C8H8)2$ |
|---|---|---|---|---|---|---|---|---|
| | DFT | RUN | DFT | RUN | DFT | no +U | DFT+U | no +U |
| $A_{1g}$ | | 212.05 | 215 | 222.0 | 219 | 216.9 | 212.5 | 217.0 |
| | | | 735 | | 709 | | | |
| | 722 | 752.5 | 776 | 750.3 | 742 | 752.4 | 750.7 | 750.2 |
| | 2968 | 3041.1 | 3069 | 3042.2 | 3122 | 3065.2 | 3072.3 | 3064.6 |
| $E_{1g}$ | | 236.2 | 237 | 241.7 | 206 | 237.3 | 231.5 | 242.8 |
| | 595 | 725.62 | 774 | 722.1 | 750 | 740.5 | 694.4 | 734.6 |
| | | | 918 | | 899 | | | |
| | | | 1439 | | 1421 | | | |
| | | 3168.7 | 3055 | 3020.7 | 3103 | 3108.8 | 3109.5 | 3064.6 |
| $E_{2g}$ | | | | | | | | 255.2 |
| | | | 235 | 263.7 | 253 | 366.2 | 362.9 | 368.4 |
| | 325 | 379.86 | 378 | 386.8 | 364 | 897.6 | 897.1 | 892.4 |
| | | 896.56 | 852 | 893.2 | 828 | | | |
| | 1127 | 1143.34 | 1194 | | 1163 | 1155.7 | 1172.9 | |
| | 1452 | 1499.72 | 1520 | 1497.8 | 1496 | 1498.8 | 1500.2 | 1492.5 |
| | 2899 | 3014.3 | 3042 | 2981.6 | 3091 | 3065.2 | 3072.3 | 3064.6 |

J. Chem. Phys., Vol. 120, No. 6 (2004)